\documentstyle[12pt]{article}
\newcommand{\beq}{\begin{equation}}   
\newcommand{\eeq}{\end{equation}}

\newcommand{\ra}{\rightarrow}

\newcommand{\gsim}{\lower.7ex\hbox{$
\;\stackrel{\textstyle>}{\sim}\;$}}
\newcommand{\lsim}{\lower.7ex\hbox{$
\;\stackrel{\textstyle<}{\sim}\;$}}

\def\lsim{\mathrel{\rlap{\lower3pt\hbox{\hskip0pt$\sim$}}
    \raise1pt\hbox{$<$}}}         
\def\gsim{\mathrel{\rlap{\lower4pt\hbox{\hskip1pt$\sim$}}
    \raise1pt\hbox{$>$}}}         

\newcommand{\aver}[1]{\langle #1\rangle}

\newcommand{\La}{\overline{\Lambda}}
\newcommand{\Lam}{\Lambda_{\rm QCD}}

\newcommand{\al}{\alpha}
\newcommand{\as}{\alpha_s}
\newcommand{\GeV}{\,\mbox{GeV}}

\newcommand{\matel}[3]{\langle #1|#2|#3\rangle}

\begin{document}
\begin{titlepage}
\renewcommand{\thefootnote}{\fnsymbol{footnote}}

\begin{center} \Large
{\bf Theoretical Physics Institute}\\
{\bf University of Minnesota}
\end{center}
\begin{flushright}
TPI-MINN-97/06-T\\
UMN-TH-1532-97\\
hep-ph/9703437
\end{flushright}
\vspace{.2cm}
\begin{center}
{ \Large
{\bf Key Distributions for Charmless Semileptonic B Decay}} 
\end{center}
\vspace*{.2cm}
\begin{center} {\Large R. David Dikeman and N.G.~Uraltsev $^*$} \\
\vspace{0.2cm}
{\it  Theoretical Physics Institute, Univ. of Minnesota,
Minneapolis, MN 55455}\\
\end{center}

\vspace*{1.2cm}

\begin{abstract}                      
We present theoretical predictions for a few phenomenologically
interesting distributions in the semileptonic $b\ra u$ decays which are
affected by Fermi motion. The perturbative effects are incorporated
at the one-loop level and appear to be very moderate. Our treatment of
Fermi motion is based directly on QCD, being  encoded in the
universal distribution function $F(x)$. The decay distributions in the
charged lepton energy, invariant mass of hadrons, hadron energy, and 
$q^2$ are given.  We note that typically about $90\%$ of all decay
events are expected to have $M_X<M_D$; this feature can be exploited in
determination of $|V_{ub}|$.

\end{abstract}

\vfill

\begin{flushleft}
\rule{2.4in}{.25mm} \\
$^*$ Permanent address: Petersburg Nuclear Physics Institute,
Gatchina, St.~Petersburg 188350 Russia
\end{flushleft}

\end{titlepage}

\section{Introduction}

The measurement 
of heavy quark decay distributions is important in extracting
key parameters of the standard model. 
Towards this end, theorists
are faced with the tough task of separating strong binding
effects, arising from strong interactions at large distances,
from the relatively simple quark-lepton Lagrangian, known at short
distances. Both perturbative corrections and little understood 
nonperturbative effects are very moderate in beauty decays if one is
interested in sufficiently inclusive, integrated characteristics \cite{buv,prl}.
In experimental studies of $b\ra u\ell \nu$ decays, attempts to separate 
$b\ra c$ transitions without explicitly identifying charmed particles,
often suggests using a narrow slice of the overall decay kinematics 
reflecting the
low invariant mass of final-state hadronic system accessible only in $b\ra
u$ decays. The strong interaction effects are magnified in this case. This fact
was realized already in early papers where the QCD-based treatment of
inclusive decay rates was elaborated. In particular, it was pointed out in
\cite{prl} that the QCD-based OPE itself leads to the 
emergence of the phenomenon of  ``Fermi motion'' of the heavy quark
inside a hadron, an effect introduced {\em ad hoc} much earlier
\cite{mod}  for the description of heavy flavor decays in
phenomenological  models. For example, the impact of Fermi motion in
simple quark models  on the invariant mass of hadrons, $M_X$, was also
addressed in \cite{barger}. The concrete manifestation of Fermi motion
in QCD has some peculiarities  \cite{prl,motion} making it somewhat
different from the simple-minded Fermi motion of quark  models. The QCD
treatment was later described in detail in a number of publications
\cite{motion,randall,wiseneub,roman,bsg}. Conceptually, Fermi motion is
similar to the leading-twist nonperturbative effects in DIS. Further
details and more extensive references can be found in \cite{bsg}.

In the present work we apply the methods of \cite{bsg} to a few experimentally
important distributions in the $b\ra u$ semileptonic decays, namely, the
distribution of the electron energy $d\Gamma/dE_{\ell}$, the invariant
hadron mass $d\Gamma/dM_X$ and the hadron energy $d\Gamma/dE_h$; we also
give a plot of $d\Gamma/dq^2$. The first three distributions are directly
affected by the Fermi motion. Our treatment is in certain aspects simplified
compared to \cite{bsg}; we will mention these elements and justification later.

\section {General Strategy -- Including Fermi Motion in Heavy Quark Decay 
Distributions}

Strong interactions have two faces in $b$-decays: one, associated with the hard
modes of the gluon fields, is to produce perturbative corrections, the
other, associated with the soft modes is to produce nonperturbative 
corrections, in
particular Fermi motion. The way to separate the hard and soft modes is
dictated by Wilson's OPE: the introduction of the hard separation scale 
$\mu$. The `identity' of the two regimes has no absolute sense and 
depends on the choice of $\mu$. 

The hadronic part of the semileptonic decays is described by five structure
functions $w_i(q_0,q^2)$, only three of which are relevant for $\ell=e$ or
$\mu$ (for definitions, see \cite{koyrakh}). The effect of Fermi motion is
encoded in the heavy quark distribution function $F(x)$; note that the
primordial momentum is normalized to
$\La=M_B-m_b$ so that we
deal with the dimensionless parameter $x$. Then 
the expression for the structure functions is written in the following way:
\beq 
w(q_0,\,q^2)\stackrel{\rm leading\; twist}{\;=\;} \int \:
w^{\rm pert}\left(q_0-\frac{x\La}{M_B} \sqrt{q_0^2-q^2},\,q^2 \right) \:
F(x)\left(1-\frac{x\La}{M_B}\frac{q_0}{\sqrt{q_0^2-q^2}} \right) \,dx
\,.
\label{d1}
\eeq
Here $w^{\rm pert}$ is a parton structure function dressed with
short-distance (perturbative) corrections. This expression coincides with those
of Ref.~\cite{motion} through leading-twist terms, summation of which 
yields the
effect of the primordial momentum distribution of the heavy quark. The form
of Eq.~(\ref{d1}) is convenient since it closely follows  the 
simple picture of
the decay of the heavy quark boosted to the velocity $x\La/M_B$ along the
direction of $\vec{q}$, with $F(x)$ determining the 
probability of the corresponding initial-state configuration.

The moments $a_i$ of $F(x)$ are given by the expectation values of local heavy
quark operators. In the adopted normalization
\beq
a_0=1\;, \qquad \qquad a_1=0\;, \qquad \qquad a_2=\frac{\mu_\pi^2}{3\La^2}\;,
\qquad \qquad a_3= -\frac{\rho_D^3}{3\La^3}
\label{d2}
\eeq
with
\beq
\mu_\pi^2\;=\; \frac{1}{2M_B} \matel{B}{\bar b (i\vec D\,)^2 b}{B}\;, 
\qquad \qquad
\rho_D^3\;=\; \frac{1}{4M_B} \matel{B}{\bar b D_\mu G_{\mu 0} b}{B}\;.
\label{d3}
\eeq
One can chose a particular functional form for $F(x)$ and adjust a few
parameters to fit the phenomenologically deduced moments. This procedure was
undertaken in \cite{bsg} where the following ansatz was suggested:
\beq
F(x)\;=\;\,\theta(1-x)\;{\rm e}\,^{cx}\,(1-x)^{\alpha}[a+b(1-x)^k]
\;.
\label{d4}
\eeq
To construct the distributions we are interested in, we thus use the 
perturbatively
corrected parton distributions $w^{\rm pert}(q_0,q^2)$, add the effect of
the nonperturbative distribution via the convolution of Eq.~(\ref{d1}), and 
calculate the
decay distributions with the full $w(q_0,q^2)$. This procedure, therefore,
amounts to averaging the perturbative decay distributions over nonrelativistic
primordial motion of the heavy quark governed by the distribution function $F$.

This approach is simplified in one important aspect. In QCD, contrary to
nonrelativistic models, $F(x)$ depends essentially on $q^2$ and on the final
state quark mass \cite{motion} even when perturbative effects are switched off.
In particular, $F(x)$ completely changes when $m_b-\sqrt{q^2} \lsim M_X$ where
$M_X$ is the invariant mass of the final hadronic system. For example, 
relations
(\ref{d2}) are modified. In the presence of gluon bremsstrahlung $M_X$ can
become large, $\gg \Lam$, even for $b\ra u$ transitions. Certain relations
between the moments of $F$ at different $q^2$ still hold,\footnote{The
concrete form of Eq.~(\ref{d1}) is chosen to preserve the lowest
moments.} and we expect that
the effects of changing $F(x)$ are insignificant in what concerns
single-differential distributions.

Another complication of the QCD description is that $F(x)$ together with $\La$
and local operators defining moments $a_i$ are normalization-point dependent
when perturbative corrections to parton distributions appear. The
perturbative structure functions are $\mu$-dependent as well; only the
convolution (\ref{d1}) is $\mu$-independent. The most noticeable effect 
that arises
is due to the running mass $m_b(\mu)$, and related variation of $\La(\mu)$
\cite{bsg}. 

\section{Fermi motion in $d\Gamma/dE_l$, $d\Gamma/dM_X$, and
$d\Gamma/dE_h$}

With the structure functions (\ref{d1}) we arrive at the following lepton
spectrum in $b\ra u\,\ell \nu$ decays:
\beq
\frac{d\Gamma}{dE_\ell}\;=\;
\int \; \frac{d\Gamma^{\rm pert}}{dE}\left( E_\ell-\frac{\bar{\Lambda} 
E_\ell}{M_B}x\right) \; F(x)\,\left(1-\frac{x\La}{M_B}\right) \; dx \;\;.
\label{d9}
\eeq
For the two other distributions, $d\Gamma/dM_X$ and $d\Gamma/dE_h$, we 
need to recall the kinematic definitions:
\beq
M_X^2\;=\; M_B^2+q^2-2M_Bq_0\;=\; [m_b^2+q^2-2m_b q_0] +
2(m_b-q_0)\La +\La^2
\label{d10}
\eeq
and
\beq
E_h\;=\; M_B-q_0\;=\; m_b-q_0 +\La\;. 
\label{d11}
\eeq
From this we get 
$$
\frac{d\Gamma}{dM_X^2}\;=\;
\frac{1}{2M_B}
\int \; dq^2 \int\; dx\; F(x)\, \left(1-\frac{x\La}{M_B}
\left(1-\frac{4q^2M_B^2}{(M_B^2+q^2-M_X^2)^2} \right)^{-1/2} \right) \; \times
$$
\beq
\frac{d^2\Gamma^{\rm pert}}{dq_0\,dq^2}\left(
\frac{M_B^2+q^2-M_X^2}{2M_B}
-\frac{x \La}{M_B}\sqrt{ \left(
\frac{M_B^2+q^2-M_X^2}{2M_B}\right)^2  -q^2} 
\,,\; q^2\right) 
\label{d12} 
\eeq 
and 
$$
\frac{d\Gamma}{dE_h}\;=\;
\int \; dq^2 \int\; dx\; F(x)\,\left(1-\frac{x\La}{M_B} \left(
1-q^2/(M_B-E_h)^2 \right)^{-1/2} \right) \; \times
$$
\beq
\frac{d^2\Gamma^{\rm pert}}{dq_0\,dq^2}\left(m_b-E_h+\La \left( 
1-\frac{x\,\sqrt{\left(M_B-E_h\right)^2-q^2}}{M_B}
\right),\; q^2\right)
\;\;.
\label{d13}
\eeq

\section{Perturbative corrections}

Perturbative corrections are parametrically enhanced in the kinematics close to
the free-quark decay. In particular, in $b\ra u$ ($b\ra s$) transitions
double log effects appear. These enhanced corrections must be summed up in the
decay $b\ra s+\gamma $ where $q^2=0$, which was done in \cite{bsg}, including
the effect of running $\as$. In semileptonic transitions the bremsstrahlung
effects are much softer since the typical configuration corresponds to a
significant invariant mass of the lepton pair. Therefore, we use instead the
exact one-loop radiative corrections calculated in Ref.~\cite{czarj}; the
running of $\as$ is thus discarded as well. It seems justified since even in
the case of $b\ra s+\gamma $ these effects -- although very important in purely
perturbative calculations including integration over low-momentum gluons, -- 
are
marginally seen when one proceeds with the Wilson's OPE where the evolution of
the coefficient functions and effective low-energy parameters is done on the
same footing. We will demonstrate also that the results are fairly insensitive
to variation of $\as$, which is another justification.

On the other hand, limiting ourselves to the one-loop fixed-$\as$ calculations,
and keeping in mind the extremely weak numerical $\mu$-dependence of the 
obtained
physical spectrum,\footnote{The residual $\mu$-dependence always remains in
theoretical calculations due to their approximate nature.} we formally put
$\mu=0$ which allows using the expressions of Ref.~\cite{czarj} literally. 
This,
however, necessitates using the perturbative one-loop value of the $b$-quark
pole mass 
\beq
\tilde m_b\;\simeq\; m_b(\mu) + \frac{16}{9}
\frac {\as}{\pi} \mu\;,
\label{d15}
\eeq
likewise a similar subtracted  value of $\mu_\pi^2$
\beq
\tilde \mu_\pi^2\;\simeq\; \mu_\pi^2(\mu) - \frac{4}{3}
\frac {\as}{\pi} \mu^2\;,
\label{d16}
\eeq
etc. (for more details, see review \cite{rev}). We use
$\as/\pi\simeq 0.1$ in the analysis; at $\mu\approx 1\GeV$ it 
corresponds to $\tilde m_b\simeq
4.82\GeV$, and $\tilde \mu_\pi^2$ being about $ 0.1 \GeV^2$ smaller than
$\mu_\pi^2$ normalized at the scale $0.5$--$1\GeV$.

Although the perturbative distributions are known to one loop only, one can
easily write them in exponentiated form (see, e.g. \cite{bsg}). This is
technically convenient since the elastic peak affected by virtual
corrections disappears and one can then deal with smooth perturbative 
distributions.
One relatively simple form of the exponentiated corrections for the 
double 
distribution was 
suggested in Eq.~(17) of Ref.~\cite{greub}. To simplify
the calculations we used this form of radiative corrections for $d\Gamma/dM_X$
and $d\Gamma/dE_h$ where double-differential distributions must be used. The
numerical difference of exponentiation is negligible after 
incorporating the effect of the primordial Fermi motion.

\section{Results and Discussion}

The results of calculation of the distributions are shown in Figs.~1--3
($d\Gamma/dE_\ell$), Fig.~4 ($d\Gamma/dM_X$) and Fig.~5 ($d\Gamma/dE_h$). Our
basic set of parameters 
is $\tilde m_b=4.82\GeV$, $\tilde\mu_\pi^2=0.2$, $0.4$ and
$0.6\GeV^2$ which correspond to 
\beq
[\al,\:a,\:b,\: c]\;=\; [1.5,\:3,\:16,\: -3.3]\,,\;\; 
[0.5,\:2,\:0.76,\: -1.75]\,\mbox{ and } [0.1,\:1.2,\:0,\: -1.1]\;,
\label{27}
\eeq
respectively, in the ansatz for $F(x)$ (we use $k=1$); the value of $\as$ is
set to $0.3$.

To illustrate dependence on $m_b$ we show also the plot for $d\Gamma/dE_\ell$
for $\tilde m_b=4.73 \GeV$ (Fig.~2). 
Figure~3 shows the dependence of the lepton
spectrum on $\as$; it is clearly rather weak.

As in the case of $b\ra s+\gamma$, the effect of the Fermi motion  -- where
present -- is more pronounced than perturbative modifications. As expected, it
shows up in a much softer way in $d\Gamma/dE_\ell$ compared to
$d\Gamma/dE_\gamma$ in $b\ra s +\gamma$. The situation with $b\ra s 
+\gamma$ is in  closer analogy with 
$d\Gamma/dM_X$, but even here the effect is suppressed due to significant
average invariant mass of the lepton pair, i.e. effectively smaller energy
release and recoil momentum.

Apart from smearing by Fermi motion of non-smooth parts of the partonic
distributions, the most significant nonperturbative effect in $d\Gamma/dM_X$ and
$d\Gamma/dE_h$ is associated with the difference $\La=M_B-m_b$ in the value of
$E_h=M_B-q_0$. This nonperturbative effect was first pointed out and
analyzed in \cite{WA}. 

The predicted shape of the distributions depends to some extent on the 
adopted form
of the distribution function  $F(x)$. We believe, nonetheless, that
possible variations are not significant as long as the same 
underlying parameters, $m_b(\mu)$ and $\mu_\pi^2(\mu)$ are employed.

A more significant effect is possible from the next-to-leading twist
operators, since the effective energy release is not very large. An
example of such effects is the chromomagnetic interaction whose effect can
be estimated \cite{motion,bsg}, for instance, as due to the final-state mass 
difference between $\pi$ and $\rho$, or due to the initial-state mass
difference for weak decays of $B$ and $B^*$ -- these effects 
are neglected in
the analysis. They can cause certain shifts of the
distributions which are predicted if smearing over the corresponding
interval is done. For example, in the distribution over $M_X^2$ the
necessary interval of smearing over $M_X^2$ constitutes, probably, about $\pm
0.2\GeV^2$. In the regular parts of the distributions such a smearing is
superfluous, of course.

For the same reason the very beginning of the distribution over $M_X$
and $E_h$ cannot be taken literally; in particular, one {\em cannot}
deduce from it the exclusive decay rate into $\pi$ or $\rho$, or
any particular resonance. In the limit $m_b\ra \infty$ the distance
between the successive resonances in $M_X$ would disappear in the scale 
of $\aver{M_X}$; for the 
actual case of $B$ it still produces a relatively coarse `grid'.

A related peculiarity of the presented plots is that they show a
nonvanishing rate for $M_X$ or $E_h$ below pion mass, which, from the OPE
viewpoint, is completely an
artefact of neglecting higher-twist effects. To get rid of this
unphysical feature one can, for example, consider {\em ad hoc} the given
distribution as the one over $M_X^2-M_\pi^2$ rather than over $M_X^2$ --
distributions modified in such a way are  formally equivalent to the 
order in  $1/m_b$ expansion we work.

The decay distribution $d\Gamma/dq^2$ is not essentially affected by
strong interactions. At maximal $q^2$ particularly close to
$m_b^2$, nevertheless, these effects blow up. It is important to
realize that it is not literally the effect of  Fermi
motion that shapes $d\Gamma/dq^2$. For example, the integral over
the large-$q^2$ domain is governed by the flavor-dependent expectation value 
of the six-quark operator and thus can be completely different in decays of
$B^\pm$ and $B^0$. The effects originating at $q^2 \ra m_b^2$ were
discussed in detail in \cite{WA}. Here we remind that studying the
difference of the decay distributions for $B^\pm$ and $B^0$ near maximal
$q^2$, at small $M_X$ or $E_h$ -- or merely in the end-point electron
spectrum -- is a way to directly measure the four-fermion expectation
values which are important for $B$ physics.

The decay distribution $d\Gamma/dq^2$ is shown in Fig.~6; except near the
kinematic boundary, the deviation from the tree-level parton shape is
very small. We obtained this distribution merely adding known
nonperturbative $1/m^2$ effects \cite{koyrakh} to the parton 
expressions calculated through order $\as$ \cite{czarj}.
\vspace*{.15cm}

An interesting observation we infer from the plots is that a significant 
fraction, $\approx 90\%$,  of the decay events is expected to have $M_X <
1.87\GeV$, i.e. lie below the charmed states. 
It is in contrast with the case of semileptonic spectrum where only a small
fraction of the decays proceed to the domain above the kinematical bound for 
$b\ra c$ transitions, and even small measurement errors lead to sizeable
bias.
For the distribution in invariant mass, 
with a lower cutoff on $M_X$ between  $1.5$ and $1.6 \GeV$ 
(to allow for a possible leaking of higher-mass states due to experimental
uncertainties), the majority of decays appears in the low-$M_X$ region
($80\%$ for $M_X< 1.6\GeV$ and about $75\%$ for $M_X< 1.5\GeV$), and can 
be reliably calculated theoretically. On the other hand, this can possibly 
be determined in experiment. This
would suggest a way for a trustworthy determination of $|V_{ub}|$ with a
relatively good accuracy.

If a measurement of the distribution over $M_X^2$ is possible, it yields
the
possibility to determine independently the hadronic parameters. In analogy to
the $b\ra s+\gamma$ decays, the center of gravity of the distribution is
sensitive to the $b$ quark mass and its width determines $\mu_\pi^2$. Figure~4
shows a reasonable sensitivity 
of the distribution over $M_X^2$ to $\mu_\pi^2$.

\section{On a Possible Improvement of Theoretical Predictions}

The presented analysis of the decay distributions is in many aspects
simplified. While the $b$ quark mass is large enough for fully integrated
characteristics, it is not the case when the detailed differential predictions
are attempted. It is not therefore clear that further refinements based on more
advanced treatment of perturbative corrections and nonperturbative effects can
yield an essential and trustworthy improvement. The first candidate for the
refinement is, obviously, inclusion of known $1/m^2$ corrections which 
are not
incorporated in the leading-twist effects summed up by the Fermi motion. The
anticipated scale of these effects was mentioned above.

Another direction for refinements is improving the perturbative 
description. The
impact of the perturbative corrections on the distributions even in $B$ 
decays
appears to be smaller than those of nonperturbative effects. In particular, we
expect a small effect from inclusion of higher-order $\as$-corrections. This 
applies,
however, only when a proper treatment of the infrared domain is done --
otherwise the  running of $\alpha_s$, typically the dominant effect 
among
the second-order corrections, apparently generates effects which blow up. 
This does not happen when one literally follows Wilson's procedure of
constructing the OPE. Experience shows that this procedure is 
necessary 
already when the second-order corrections in $b$ decays are addressed - 
the procedure enforces the safeguard from inconvenience of 
the miracleous compensation of significant effects coming from
different sources \cite{upset,blmope}.

While introduction of the IR cutoff in calculating the perturbative 
coefficient
functions is more or less straightforward in usual perturbative calculations,
this is not so simple when exponentiation of soft/collinear effects is
employed. A method applicable to these problems was described in \cite{bsg}; it
combines a few essential elements: reproducing exact one-loop and 
(all-order) 
BLM-improved answer and, simultaneously, proper exponentiation of singular 
double-log corrections. This feasible way 
is expected to yield more than enough
accuracy in evaluating perturbative corrections.

While such a treatment is necessary when one intends to determine underlying
parameters of the heavy quark expansion like $m_b$ and $\mu_\pi^2$ with the 
accuracy when 
their scale-dependence is essential, we anticipate a small overall impact of
the higher-order perturbative 
corrections on various hadronic characteristics involved, for
example, in measuring $|V_{ub}|$, well below the effect of the leading
nonperturbative phenomena.

\section{Outlook}

One of the most important practical applications of the analysis of the
$b\ra u\,\ell\nu$ distributions is the extraction of $|V_{ub}|$. As was
realized long ago \cite{Ds}, the theoretically cleanest way to
determine it is from the total $b\ra u\,\ell \nu$ width. In
particular \cite{upset},
\beq
|V_{ub}|\; \simeq\; 
0.00415\left(\frac{{\rm BR}(B\rightarrow X_u\ell\nu)}{0.0016}
\right)^{\frac{1}{2}}\left(\frac{1.55\,\rm
ps}{\tau_B}\right)^{\frac{1}{2}}
\label{d40}
\eeq
where theoretical uncertainties lie well below the level which is 
experimentally relevant now and in the near future. However,  a completely 
model-independent
experimental measurement of such a width is embryonic at present.

Another extreme way, suggested long ago, was to consider decay
events with $E_\ell \gsim (M_B^2-M_D^2)/2M_B = 2.31\GeV$, where charm
decays cannot contribute. The decay rate in this too narrow slice of
kinematics, on the other hand, depends on poorly known details of
strong dynamics. Within our calculation this fraction of all decay events 
does not vary too significantly; nevertheless, more work is needed to
put this observation on a more firmly grounded quantitative basis.

There are many intermediate options whose advantages mainly depend on
experiment. The larger the kinematic domain, the more it encompasses the
parton-level kinematics, the better, in general, is the theoretical
description of such an inclusive width. We found that the distribution
over $M_X$ studied in experiment is rather promising in this respect. A
similar suggestion was made to look at the distribution over $E_h$
\cite{greub}. It is clear that an accurate determination of $|V_{ub}|$
requires a detailed analysis  of the kinematics of the decay events,
which would ultimately lead to determination of just the
double-differential distributions over $q_0$ and $q^2$. Which particular
integral appears to be most suitable for extracting  $|V_{ub}|$ will
eventually depend on the available experimental technique.

\vspace*{0.5cm}

\noindent
{\bf ACKNOWLEDGMENTS:} \hspace{.4em} 
We are grateful to P.~Henrard and his ALEPH collaborators for suggesting
to us the extension of our $b\ra s+\gamma$ considerations to the
calculation of the semileptonic $b\ra u$ distributions, useful
discussions, critical  comments on the results and incorporating them in
the analysis of the ALEPH data \cite{rosnet}.  We are thankful to
I.~Bigi and M.~Shifman for stimulating comments and their  encouraging
interest.  N.U. thanks the CERN Theory Division for kind hospitality
during the period when the main calculations of this work  were done.
This work was supported in part by DOE under the grant number
DE-FG02-94ER40823 and by NSF under the grant number PHY 92-13313.

\newpage
\begin{figure} 
\vspace{8cm}
\includegraphics{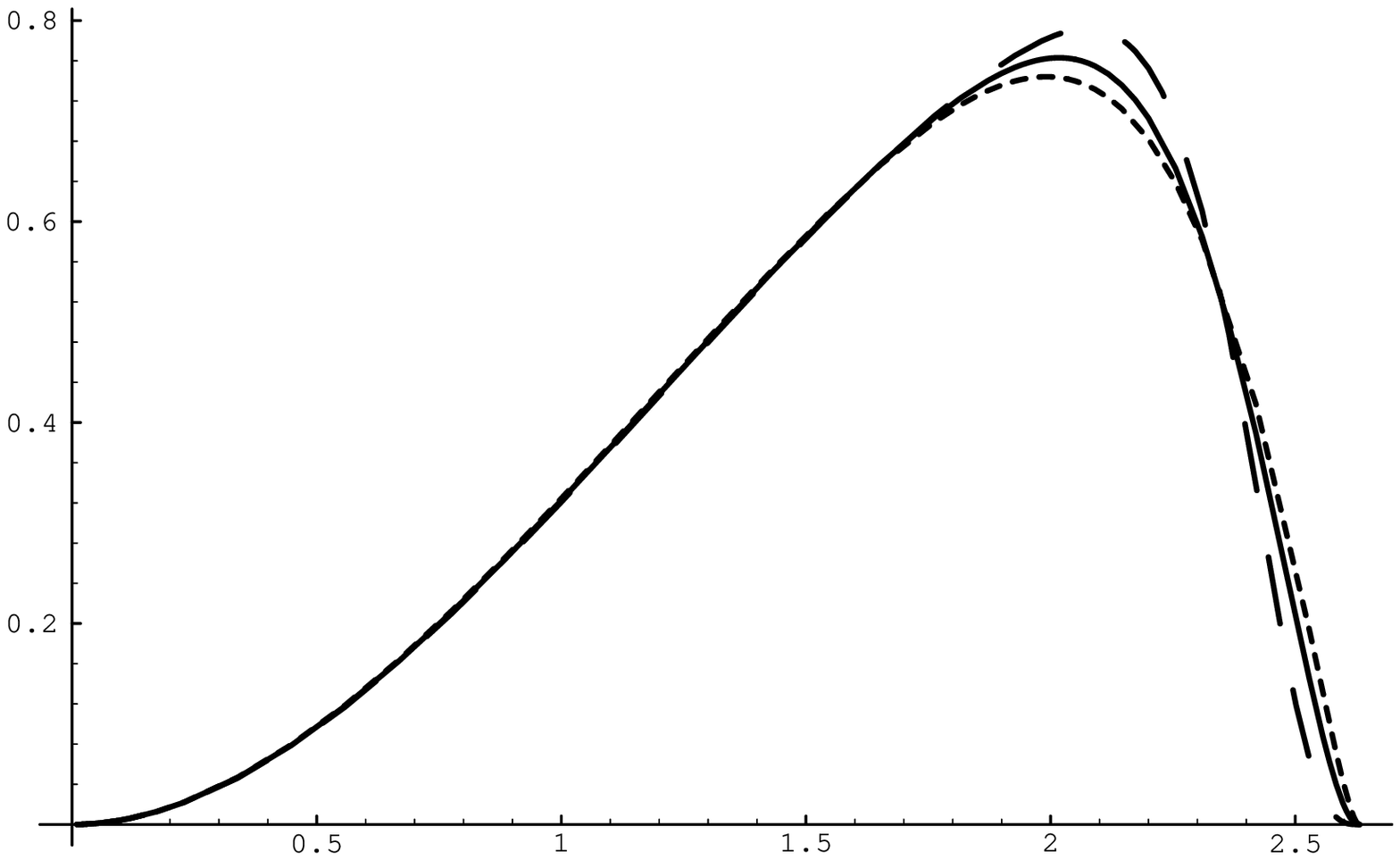}
\caption{
The electron energy distribution $d\Gamma_{\rm sl}(b\ra u)/ dE_l$,
arbitrary units.
Long-dashed, solid, and short-dashed
lines correspond to $\tilde \mu_\pi^2 =0.2$, $0.4$
and $0.6\GeV^2$. The $b$-quark mass $\tilde m_b=4.82\GeV$, $\as=0.3$.}
\end{figure}

\begin{figure}
\vspace{8cm}
\includegraphics{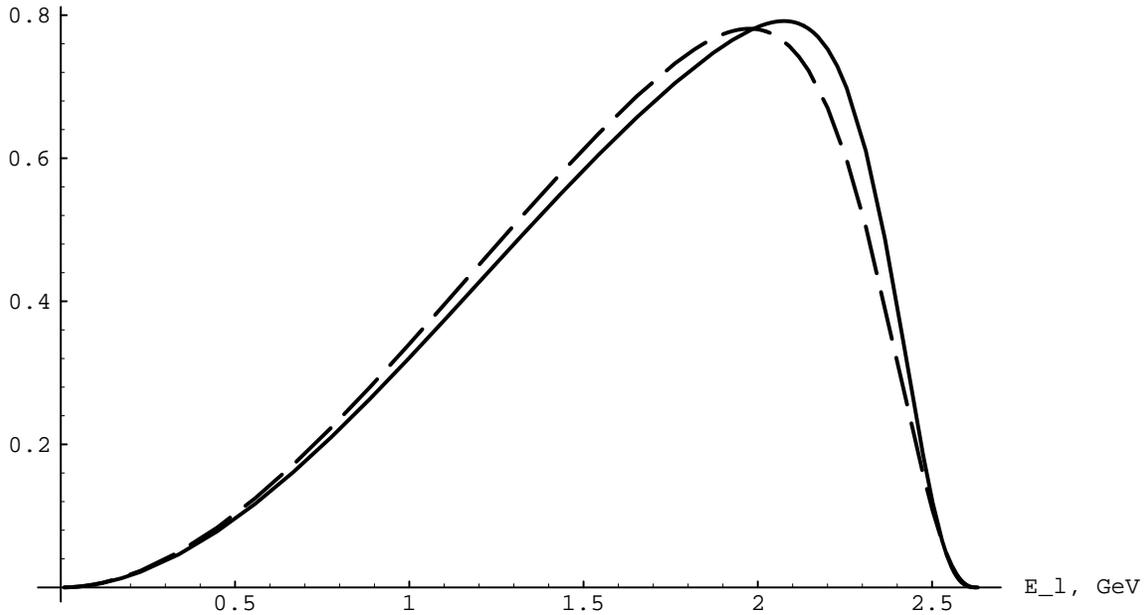}
\caption{
Dependence of $1/\Gamma_{\rm sl}(b\ra u)\;d\Gamma/ dE_l$ on
$m_b$. The solid line corresponds to $m_b=4.82\GeV$, and the dashed line
is for $m_b=4.72\GeV$,  while $\tilde\mu_\pi^2 = 0.4\GeV^2$ is kept
fixed.
}
\end{figure}
\newpage

\begin{figure}
\vspace{8cm}
\includegraphics{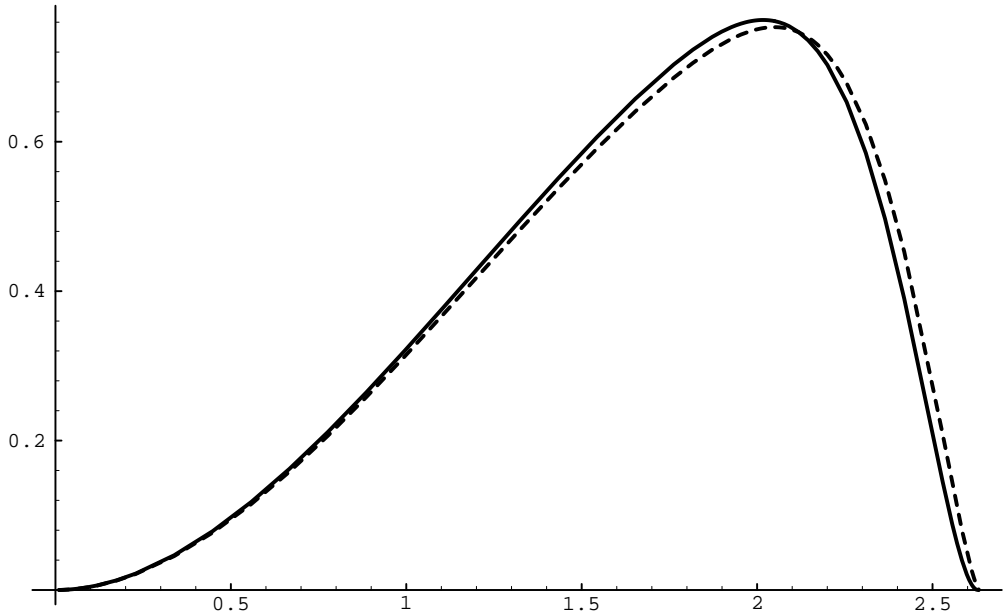}
\caption{
Effect of radiative corrections on $d\Gamma/dE_l$:
solid line shows $\as=0.3$ and dashed line is $\as=0$.
Here
$\tilde m_b=4.82\GeV$ and $\tilde \mu_\pi^2 = 0.4\GeV^2$.
}
\end{figure}

\begin{figure}
\vspace{8cm}
\includegraphics{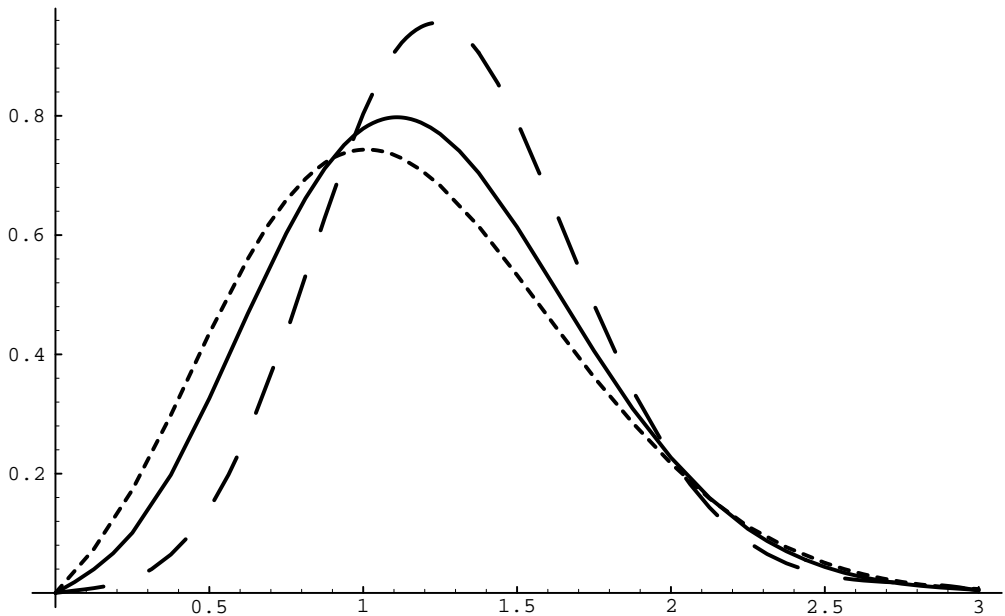}
\caption{
The invariant mass distribution $d\Gamma/dM_x$; long-dashed, solid,
and short-dashed lines correspond to $\tilde \mu_\pi^2 =0.2$, $0.4$
and $0.6\GeV^2$. The $b$-quark mass $\tilde m_b=4.82\GeV$, $\as=0.3$. All
distributions are normalized to the same total width.}
\end{figure}

\newpage
\begin{figure}
\vspace{8cm}
\includegraphics{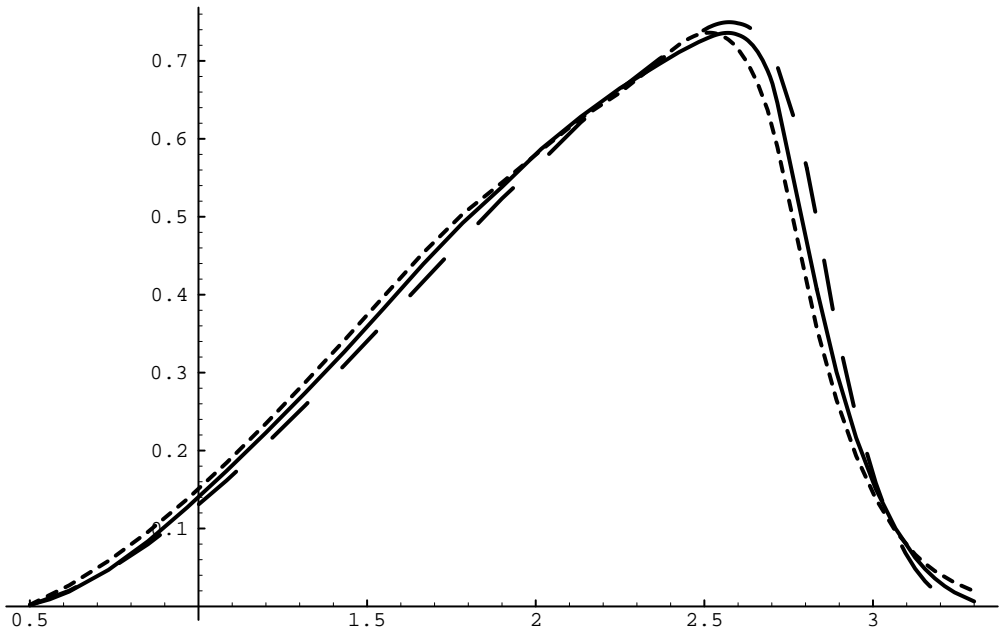}
\caption{
The hadronic energy distribution $d\Gamma/dE_h$ in the same
setting as in Figs.~1 and~4.
}
\end{figure}

\begin{figure}
\vspace{8cm}
\includegraphics{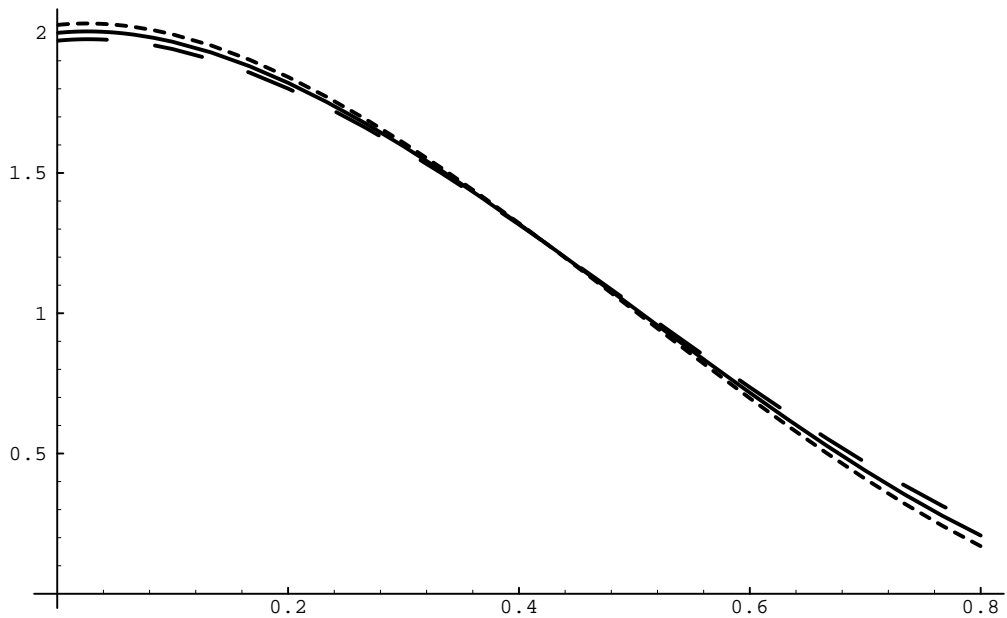}
\caption{
The $q^2$-distribution $d\Gamma/dq^2$ is made with the same
conventions as for Figs.~1 and 4.}
\end{figure}

\end{document}